# Female Agency and its Implications on Mental and Physical Health: Evidence from the city of Dhaka

*Authors:*

**1. Upasak Das** (corresponding author)

Global Development Institute, University of Manchester
Center for Social Norms and Behavior Dynamics, University of Pennsylvania
ORCID: 0000-0002-4371-0139

*Email*: upasak.das@manchester.ac.uk

**2. Gindo Tampubolon**

Global Development Institute, University of Manchester
ORCID: 0000-0002-9081-2349

*Email:* gindo.tampubolon@manchester.ac.uk

**Declarations of interest:** None

**Acknowledgements**

The authors would like to thank the participants of a webinar based on the paper organized by the Global Development Institute, University of Manchester.

**Conflict of Interest Statement**

We would like to confirm that there are no known conflicts of interest associated with this work and there has been no significant financial support that could have influenced its outcome.

**Funding disclosure**

This research received no external funding.

**Abstract**

*Women agency defined as the ability to conceive of purposeful plan and to carry out action consistent with such a plan can play an important role in determining health status. Using a sample of 1259 female respondents from the city of Dhaka in Bangladesh, this paper explores how women agency relates to their physical and mental health. The findings indicate women with high agency to experience significantly lesser mental distress on average. Counter-intuitively, these women are more likely to report poor physical health. As an explanation, we propose purposeful action among women with high agency as a potential reason, wherein they conceive purpose in the future and formulate action that is feasible today. Hence, these women prefer to report illness and get the required treatment to ensure better future health. This illuminates our understanding of sustainable development and emphasises the critical role of women's agency for sustainable human development.*

# 1. INTRODUCTION

Research on health and one's well-being has indicated that both are linked to how well the person concerned controls his/her own life. In a patriarchal context with entrenched gender norms, women empowerment and the autonomy she enjoys within households can be related to her physical and mental well-being. This paper, using a sample of 1259 female respondents from the city of Dhaka in Bangladesh examines how mental health of a woman is associated with her agency. In addition we also study if women with higher agency are able to leverage their better bargaining power within the household and community and improve their physical health.

Scholars have conceptualized women agency as their ability to conceive of purposeful plan and to carry out action consistent with such a plan. Therefore, it not only includes intrinsic qualities like having critical and positive opinion but also instrumental ones like taking decisions (Kabeer, 1999; Mosedale 2005). Because agency is often facilitated by better access to income, education and social status among others, it can potentially have important implications on their mental as well as physical health. In this respect, Richardson et al. (2019) studied the association between agency and mental distress among rural women from the Indian state of Rajasthan. Our study complements this by examining the similar issues in the context of urban women from Dhaka. The contextual difference between the two regions, however is rural Rajasthan fares very poorly in terms of gender parity while the Dhaka city performs better in this front largely driven by female participation in the livelihood generating process (Salway et al. 2003; Bose et al. 2009; Government of India, 2018). Nevertheless, historically both regions have strong patriarchal norms biased against the women.

One of the key contributions of the paper is to assess female agency and its relationship with both the aspects of their health- the physical and mental health. Alongside physical health, the problem of mental health assumes importance as public health challenge because of its high prevalence along with a number of direct and indirect detrimental effects. This is especially pertinent for females as it has been consistently documented that prevalence of anxiety and depression is greater among women than among men (Seedat et al., 2009; Steel et al., 2014). As found in a recent study by World Health Organization (WHO), about 10 percent of pregnant women are victims of problems related to mental health and depression which is observed in about 13 percent of women who have just given birth.[1] This problem seems systematically higher among females from developing countries as the prevalence of mental health problems is found to be close to 50 percent higher than in developed countries. As put forward by Bendini and Dinarte (2020), close to 16 percent of women from these countries experience depression during the period of pregnancy and post childbirth, this figure goes up to about 20 percent. Notably, given the dearth of availability of quality data on maternal health problems especially in developing countries, the figures on the prevalence of mental depression is likely to be an under-estimate and represent lower bound of a much larger problem (Parsons et al 2012).

Importantly, Bangladesh has been reported to have high burden of mental health disorders. A survey conducted nationally in the country on mental distress among adults found more than 16% to suffer from mental depression and women are found to have systematically higher prevalence (Firoz et al. 2006). High prevalence of mental disorder has also been reported in other studies (Hossain et al. 2014; Wadood et al. 2020). Notably, studies have found mental anxiety among females to have a significant role in household economy and also in child and elderly care (Islam et al. 2018). For instance, Bendini and Dinarte (2020) document depression among mothers

---

[1] https://www.who.int/teams/mental-health-and-substance-use/maternal-mental-health (accessed on December 2, 2021)

adversely affecting learning outcomes of their children. It has also been documented that an increase in income levels is associated with a reduction in mental depression levels and that negative impact of exposure to idiosyncratic shocks associated with poverty can potentially be intensified by higher mental disorder and depression (Tampubolon and Hanandita, 2014). It is in these contexts that female physical health also assumes significance. In sum, better health for women is directly related to higher productivity in the labour market in addition to it ensuring better education and health for children as well. This necessitates the need to study physical illness along with mental distress.

Our findings throw up interesting inferences that has important contextual relevance. Women with higher agency especially those who are more likely to speak up in public suffer from lesser mental depression. This is in contrast to Richardson et al. (2019) who find women who are more likely to speak up in public have higher chances of suffering from distress. The authors argue it is easier for females who are comfortable in speaking up in public to reject the highly entrenched gendered norms which restrict women to do so. Because the context of study is rural Rajasthan, the norms are highly patriarchal and hence this can potentially cause a pushback from the community and the resulting social sanctions may lead to higher mental distress among women. In the context of Dhaka, with more women working outside, it is possible that the ability to speak up in public can improve their mental health. Nevertheless, we also find women with higher agency to have disproportionately higher probability of suffering from physical illness. The argument that we put forward is that these women are capable of conceiving purpose in the future and formulating action that is feasible today. Accordingly, they report of being ill presently so that necessary actions can be taken to ensure better health in the future.

The importance of our study lies on the fact that the implications of women agency are, to a large extent, dependent on the context. Women's agency does not invariably lead to better mental health- in certain contexts, it does while in others, it may not. One possible dimension to this is assessing how exercising the agency for a woman interacts with the societal norms on women's role and the likely response of other groups in the community. Our results here open up a promising theoretical and empirical inquiry into this topic. In addition, it also paves way for future research on how women with higher levels of agency give emphasis to physical health and future well-being, which can ensure formulating better actions in the present.

## 2. DATA

Data used in the paper come directly from the DIGNITY (Dhaka low Income area GeNder, Inclusion and poverTY) survey, that was specifically conducted in 2018 to shed more light on issues including poverty, empowerment, and livelihood in the city of Dhaka from Bangladesh (The World Bank Group, 2018). The survey is representative of low-income areas of the Dhaka City Corporations including slums that cover the northern as well as the southern parts and an additional low-income site from the Greater Dhaka Statistical Metropolitan Area. The survey strategy follows a two-stage stratification design, wherein in the first stage the Primary Sampling Units are selected using Probability Proportional to Size and 20 households were selected for interviewing from each sampling units.[2]

---

[2] The data is available on request from https://microdata.worldbank.org/index.php/catalog/3635/related-materials.

The survey collected information on a number of issues related to female economic empowerment and livelihood. Apart from the demographic and socioeconomic characteristics, it gathered information on detailed work history, time use, and perceptions about work. Importantly it administered questions on a number of indicators that pertain to her autonomy and agency along with mental anxiety and distress. In addition, it also collected data on her health status. The survey instrument mainly has two modules: the more traditional household module where the head of household is interviewed to collect the basic information on the household and then the individual module, where from each household two respondents are interviewed separately. Specifically, the main couple from each household is selected for the interview. Notably, to minimize response bias, one male and one female interviewer were deployed for each household, such that the gender of the respondents matched with that of the surveyors.

Figure 1 gives the locations of the survey areas. As one may observe, the survey areas are well spread across the city of Dhaka. The total sample comprises 1300 households from which 2376 individuals have been surveyed. Out of them, 1259 are females and the remaining 1117 are males.

[Figure 1 here]

## 3. VARIABLES

This paper examines the relationship between women agency and their mental as well as physical health indicators. Accordingly, we have two sets of outcome indicators: the first set pertains to mental disorder and distress and the second set pertains to physical health. The following variables have been used to estimate the former: *In the last 7 days*

1. *were you disturbed by things that don't normally bother you?*
2. *did you have trouble concentrating on what you were doing?*
3. *did you feel depressed?*
4. *did you feel that everything you did was a burden?*
5. *did you feel afraid?*
6. *was your sleep restless?*
7. *did you feel lonely?*
8. *did you feel like not getting up in the morning?*

These indicators of mental depression from the Centre for Epidemiologic Depression Scale (CESD) are widely used in the context of developed and developing countries (Tampubolon and Hanandita, 2014; Ohrnberger et al. 2017; Tampubolon and Maharani, 2017). The response was collected in terms of the number of days, where 0 indicated that the respondent did not have these anxieties and 7 indicated that she faced the above in all the days of the week. To get an overall mental anxiety score, we first standardized each of the above indicators and calculated the sum total of these standardized indicators. The final mental anxiety score is the standardized value of this sum total and allows us to interpret the inference in terms of standard deviations.

To measure physical health condition, we use the following question: Have you suffered from ill-health that has limited your ability to undertake your normal daily activities in the last 30 days? In addition, we also use the number of days of illness for those respondents who reported ill health on the above question.

Our main variable of interest is women agency. To calculate it, we use a set of literature on women agency in different context and find the specific questions in the questionnaire that captures the essence of different dimension of women autonomy. Accordingly, the following questions are used to construct agency:

1. *Do you usually cover your head or wear a burkha when walking in the streets outside your community?*
2. *Do you ask permission from anyone in your household to travel outside your community by yourself?*
3. *Do you feel that a woman like yourself can generally change things in the community where you live if she wants to?*
4. *Do you feel comfortable speaking up in public to protest the misbehavior of authorities or elected officials?*
5. *Do you feel comfortable speaking up in public to help decide on infrastructure (like tube wells/community toilets, walkways/roads, common kitchen) to be built in your community?*
6. *Do you alone have any money you can decide what to spend on?*

The question on burkha was asked with three options: No covering, head covering and wear burkha. We grouped the first two options and assigned it with a value of 1 and the last response was given a value of 0. The second question on permission to travel outside by herself was asked with two options: Yes or No. Those who responded "Yes" were assigned a value of 1 and 0 was assigned for those who said "No". On similar lines, we also code the third and the sixth questions.

On the fourth and fifth question, the respondent is assigned the value of 1 if she says she can speak up comfortably or with difficulty and 0 if she is unable to.

As one can observe, through these six questions, we tapped into different dimensions of agency including freedom to go against social norms (not wearing burqa outside the community); freedom of movement (asking for permission to travel outside the community by herself); self-confidence (can change things in the community where she lives); engagement with public speaking, individual leadership and influence (speaking up on misbehaviour of authorities or elected officials and community infrastructure) and decision making (decision of what the money can be spent on). Prior theoretical and analytic literature on women empowerment and agency guided us to select these variables from the questionnaire. As an instance, the theoretical work on the Capabilities Approach defines agency as the ability to identify goals and then act upon them (Sen, 1985; Kabeer, 1999; Ibrahim and Alkire, 2007; Iqbal et al., 2020). In addition, we also use Malapit et al. (2019) and Richardson et al. (2019) to decide these variables and use them to generate women agency scores. To calculate these scores, we use the same methodology that has been used to generate those on mental anxiety. Accordingly, for each of the six indicators, we generate a standardized value and then take a sum total of these. Finally the women agency score is calculated by standardizing this value. Of note is the fact that this aggregation method has been used by recent papers on women autonomy (Heath and Tan, 2020).

We control for an extensive set of covariates in our regressions that are likely to be confounded with mental and physical health of an individual. Firstly, age, education and marital status of the respondent is incorporated in the model as they can directly influence mental anxiety and health conditions. We also control for whether the respondent is the head of the household along with the gender of household head. Household heads may have systematically higher household responsibility which may affect her health and mental conditions adversely. Further, female headed households are likely to economically poorer and victims of social deprivation that acts through gender. This can affect the physical health and mental stress levels. Economic condition of the household is accounted for by the total consumption expenditure on items that include milk, milk products, meat, fish and eggs. Further the regression also accounts for whether the household roof is cemented and it uses Liquefied Petroleum Gas (LPG) for cooking purpose. Since total consumption expenditure is dependent on the household size, we incorporate it in the model as well. It is possible household rents can be the cause of mental stress and accordingly we also control for whether the dwelling of the house is owned or rent free. Apart from this, the number of mobiles phones that the household members have is included in the regression model along with the educational level of the respondent's parents, which to some extent indicates early childhood health that in turn is related to the current health status. Finally, we account for dowry, which remains as one of the key predictors of female bargaining power within the household, through a binary variable that takes the value of 1 if the respondent reports of having brought assets from her parent's home and 0 otherwise. Other covariates include marital status and education of the respondent.

## 4. ESTIMATION STRATEGY

As indicated earlier, we examine the implications of women agency on indicators of mental health and physical illness. Accordingly, our main variable of interest is the composite score indicating agency, which forms our main explanatory variable. The composite score on mental health and illness are the outcome variables of interest. To examine the relationship between the two, we use the following regression:

$$Y_{ih} = \alpha + \beta X_{ih} + \gamma M_i + \pi H_j + \vartheta_i \qquad (1)$$

Here $Y_{ih}$ is the outcome variable for woman $i$, from household $h$. $X_{ih}$ is the main variable of interest, which is agency of the woman. $M_i$ and $H_j$ represent the set of individual and household level covariates incorporated in the model as discussed respectively. $\vartheta_i$ is the error term and assumed to follow a standard normal distribution, $N \sim (0,1)$. When we regress the composite scores of mental health, which is continuous in nature, we make use of an Ordinary Least Squares (OLS) regression. For regression of binary indicators (whether the woman suffered from physical illness or not), we use a standard probit regression. To estimate the determinants of number of days of illness, which is left censored and observable only for those who reported of being ill in the last 30 days prior to the survey, we make use of tobit regression estimation. The standard errors are clustered at the level of neighbourhood.

## 5. RESULTS

*5.1 Descriptive statistics*

Table 1 presents the descriptive statistics of the dependent variables along with the main variables of interest. For the variables, which are continuous in nature, we provide the mean and standard deviation along with the maximum and minimum value. For the binary variables, the proportion in the sample of respondents is given. We find that the average age of the female respondents in our sample is about 34 years with an average household size of about 4 members. More than 52% of them work outside with about 85% being currently married and 62% educated. More than 51% of the females indicated that they never wear a burkha while going outside. On an average, more than 60% reported that they never have to take permission to step outside of their community and about 43% of them reported that they can speak up on infrastructure.

Our sample of female respondents report that on an average out of seven days, they feel lonely on 1 day and on 1.6 days, they feel depressed. On 1.4 days, they feel disturbed and have restless sleep on more than 1.2 days on an average out of seven days. About 39% reported of being ill in the last 30 days and they remain ill for about 7 days in these one month.

[Table 1 here]

Further, to understand the relationship between women agency and indicators of mental and physical health, we first make use of the kernel weighted local polynomial smoothened regression plots along with the 95% confidence interval that makes no restrictive parametric assumptions (Fan and Gijbels, 2018). It allows us to observe the how one variable is related to another over the whole distribution and assumes no functional form. The figures given in appendix B1-B3 indicate no clear relationship of mental depression score and days of illness with female agency though a positive relationship is found for probability of reporting illness.

[Figure 2 here]
[Figure 3 here]
[Figure 4 here]

*5.2 Regression results*

As elucidated in equation (1), we run the regressions to study the extent to which women agency predicts their mental distress and their reported physical health. Figure 5 gives the regression estimates of indicators of women agency on indicators of mental distress. We present the marginal effects from regressions of mental health separately with each of the six indicators of agency and then one with the composite agency score. The findings indicate a negative association between female agency and mental depression scores, which is statistically significant at 5% level. So as one would expect, women with higher agency are more likely to be less mentally depressed. One standard deviation increase in female agency is found to be associated with about 0.08 standard deviation (CI: 0.142-0.012) fall in mental depression.

[Figure 5 here]

The regression results also indicate that the key variable on women agency that seems to predict mental distress for women is the ability to speak up on community infrastructure. It is possible that women who can speak up on community infrastructure are more likely to have better ties with the community and have widespread social network, which may help in reducing her mental distress. Notably, as mentioned this is opposite to what Richardson et al. (2019) documented where they observed women who reported of greater comfort with speaking up in public had higher levels of mental distress. While they acknowledge this might be counter-intuitive, they argue females who are more comfortable speaking up in public are able to reject the norms that restrict women speaking out publicly, which are likely to be male dominated. As a result the community may turn hostile towards them, and then this may lead to higher mental distress. Importantly, while Richardson et al. (2019) found this in the context of rural Rajasthan, we found the contrasting inferences in a highly urban setting. Our data indicates close to 53% of the woman in the working age cohort 15 to 65 years participate in the labour market with more than 43% of them reporting that they are able to speak up on infrastructure. Therefore, it is likely that social norms in our context is potentially less male dominated in comparison to that in rural Rajasthan because of which we get an expected negative relationship.

Next we examine the implications of agency on physical illness. Accordingly, we run a logistic regression since our outcome variable is a binary with respondents who reported of being ill in the last 30 days as 1 and 0 otherwise. Figure 6 presents the odds ratio from the regressions that adjust for other relevant covariates, similarly like that from Table 1. As one may observe, female agency score is found to be positively related to the reported probability of being ill. In terms of the effect size, we find that one standard deviation increase in female agency is associated with a 1.18 times (CI: 1.004-1.390) higher odds of reporting illness. With respect to the individual indicators of agency, females who do not wear a burkha along with those who are able to speak up in public protest are more likely to suffer from illness with the former having 1.6 times (CI: 1.19-2.22) and the latter having 1.53 times (CI: 1.13-2.06) greater odds of being ill.

[Figure 6 here]

Further, we consider the days of illness in the last 30 days for those respondents who reported of being ill. This arguably captures the extent of illness. We examine the implication of agency on the extent. Because extent of illness is only observable for those who reported of being ill and not for others, we make use of a tobit regression for the non-random sample of ill respondents. Figure 7 presents the regression that predicts how female agency levels are related to the days of illness. The findings, as in the earlier case reveal a significant and positive relationship between the two. We find one standard deviation increase in female agency is associated with an increase in the number of days of illness by nearly one day (CI: 0.06-1.81). The two indicators of agency: not wearing a burkha and speaking up in public protest, that are found to significantly predict the probability of being ill also are found to be pertinent in predicting the days of illness. Women who reported of not wearing burkha while stepping outside their community are associated on average with 2 more days of illness (CI: 0.32-3.85) while those who can speak up during public protests are likely to suffer additionally from 2.5 days (CI: 1.02-3.95) on average in comparison to those who do not, even after accounting for the possible confounders. In addition to these two variables, we find women who can speak up on infrastructure are found to suffer for 1.7 days more (CI: 0.02-3.39) when compared to the others and this difference is found to be statistically significant at 5% level of significance.

[Figure 7 here]

Notably, instead of standardized score of female agency, we also use Principal Component Analysis (PCA) to estimate the same. To ensure that our findings are robust to the choice of aggregation method we use to estimate agency, we run the same regressions with the PCA scores as our main variable of interest. The findings, which are presented in online supplementary table A4 indicate that our inferences remain same and statistically significant at 5% level. To sum up, we find females with higher agency to be less likely to suffer from mental depression but more likely to suffer from physical illness and that too for higher number of days. This seems counter-intuitive as women with higher agency are expected to have more control over issues that matter to them. Accordingly, literature has documented significantly positive relationship between the two (Mabsout, 2011; Ross et al. 2015). Our results however seem into indicate exactly the contrary.

What are the possible reasons for this contradiction? It is likely that women who have relatively higher agency may also have multiple household responsibilities along with potentially higher engagement within the community and outside. If this is indeed the case, the counter-intuitive result is confounded by the higher engagement in more activities. This directly comes from a set of literature that has documented higher household welfare expenditure especially towards children when the women within the household are more empowered (Behrman et al., 1999; Lam and Duryea, 1999). Research also indicates that the amount of time men spend on household work is not significantly different with the employment status of the wives (Pleck, 1985). Further, the societal expectation from a wife, independent of her employment status or agency, often include contributions to the career of her husband through support given with the household chores (Papanek, 1973). In addition, the amount of time dedicated for outside work might actually be higher for empowered women. Accordingly, the added household, labour market and societal responsibilities among women with higher agency may contribute (may be the true contributor) to their worse health that is found in the regression results.

To check this possibility, we make use of the information on time use that the survey provides for all the sampled respondents. As is the case with most of the surveys on time use, the DIGNITY survey also records information about daily activities from the respondents at every 5 minute interval in the last 24 hours prior to the survey starting from 4 am in the morning the prior day and finishing 3.59 am the current day. With this data, we calculate the proportion of time in a day which is given for household domestic works, caring for children and elders and working outside for every respondent along with that on eating/sleeping. Next, we use local polynomial regression on these over the standardized agency score for all females to examine if women with higher agency have systematic difference in the time use patterns (Figure 8). As one can expect, we find increasing relationship of time dedicated to labour related activities with higher agency. The plots for time dedicated to domestic work and caring is negatively slopped though the slope is low potentially signifying the low substitutability of domestic works for women especially when the social norms are gendered. Importantly, higher labour hours for women with higher agency can then potentially affect physical health.

[Figure 8 here]

To find if the differential time use pattern for women with high agency is driving poor health, we control for these time use patterns as well in the regression. Accordingly, we run the same regression of estimate the probability of being ill and the number of days of illness with the following additional covariates: proportion of time given for household domestic works, caring for children and elders and working as employed or own business for every respondent along with that on eating/ sleeping. The results, which are given in the online supplementary table A5 indicate no discernible changes in the marginal effects: women with higher agency are significantly more likely to report illness and even suffer from more days of illness. This makes us rule out the possibility of differential time use patterns affecting physical health for these women.

The next hypothesis we propose is related to how women with higher agency discount their future. We argue if they are able to regard themselves in their future, which is highly likely in purposeful action, they can also find it in their interest to maintain their health in the current period to derive future benefits. In other words, to ensure better health in the future, it is likely that women with higher agency would call themselves ill today and get the necessary treatment and hence prevent damage to their health in the future.

On this, the question posed during the survey on physical illness becomes important. The survey asks "Have you suffered from ill-health that has limited your ability to undertake your normal daily activities in the last 30 days?", which we use as an indicator for physical illness. It is possible that this question elicits minor illness experience only, thus cannot throw light on decisions to call ill today in order to secure future health benefits. The experience elicited by this question, being about minor illness, cannot be construed as anything related to the future. Nevertheless, one may argue because women with higher agency have different daily time use patterns and put in higher labour time on average in comparison to their peers with low agency (figure 7 above), due to the question eliciting only minor illness, the daily activities for the former may get affect while not adversely affecting the latter, who systematically give lesser time in outside work. For example, mild headache in women with higher agency may limit her ability to undertake outside labour work but with similar ailment, women with lower agency may not be able compromise her daily activities which mainly comprise household chores including caring for children and the elderly. If this is the case, one may argue that our main findings are driven by the exact wording of the question and not because of how females with higher agency regard their future health.

Unfortunately, the DIGINITY survey does not allow us to directly check this. Instead, we take representative data from the urban part of India that allows us to check the same using a similar question but worded differently. In particular, we use the 71st round of survey conducted by the National Sample Survey Organization (NSSO) on health consumption in 2014. The survey collects data on 333,104 individuals from 65,932 households that comprise of 36,480 households from rural and 29,452 from urban. For our study, we use data for women above 14 years of age (as is the case in DIGNITY survey) from the urban parts. Because the survey does not gather information on particular indicators of agency, we use data on their education which studies show is among a major predictor of agency (Duflo, 2012).

Further, the survey gathers information on physical health indicators that include any chronic ailments and incidence of hospitalization in the past year prior to the survey. In particular, the wording of the question on the chronic ailment is as follows: during the last 365 days, whether the household member suffered from any chronic ailment. On the question on hospitalization, the wording is: during the last 365 days, whether the household member was hospitalized. As one can observe, the wording of the questions in both these cases is dissimilar to the one asked in the DIGNITY survey. While the latter asked about ailments that led to disruptions in normal activity of the respondent, the former asked about ailments directly that included a question on hospitalization which is objective in nature. Accordingly, we examine whether we find similar inferences using data from women in metropolitan cities of India with questions of physical health asked differently.

We use the above two questions on physical health to create two binary outcome variables. If the woman reports chronic ailment in the last 365 days, it is coded as 1 and 0 otherwise. Similarly, if the woman is hospitalized at least once in the last 365 days, the corresponding variable is coded 1 and 0 otherwise. Our main variable of interest is that on education, wherein we create three mutually exclusive categories: illiterate, schooling below primary level and above primary level. Our regression model consists of independent variables that include age, whether there is a latrine in the household and whether the household uses modern cooking fuel (Liquefied Petroleum

Gas, natural gas, biogas, electricity) as defined by Malla and Timilsina, (2014). In addition, we also include social group or caste dummies (Upper caste as reference; Scheduled Caste; Scheduled Tribes and Other Backward Castes) because caste is among the most common correlate of social and economic deprivation across all human development dimensions including employment, education and health among others (Sundaram and Tendulkar, 2013). Further we include religion dummies along with state fixed effects in the regression model. Hence inclusion of these variables ensures we control for the possible individual and household level social and economic factors that can confound our inferences. Along with that, the state dummies ensure we account for the state level heterogeneous factors including health state level health infrastructure that can bias our regression results.

Figure 9 presents the marginal effects from the probit regression for the four educational dummy variables that we hypothesize to be a close proxy for agency. The findings indicate a significant rise in the likelihood of being chronically ill or being hospitalized with education. We observe an average of 0.094 (CI: 0.089-0.098) probability of being chronically ill for the urban females who are illiterate, which rises to about 0.11 (CI: 0.1-0.12) for women who are below primary educated. Similarly the chances of being hospitalized in the last 365 days prior to the survey for the former is about 0.25 (CI: 0.24-0.26) which increases to 0.27 (CI: 0.256-0.86) for the latter. This clearly indicates higher likelihood of being ill and even hospitalized for women who are more educated and are more likely to have higher agency.

[Figure 9 here]

Given these findings which hold in a different context in comparison to Dhaka, we reiterate our argument on women with higher agency who potentially value the future more and hence find it in their interest to act now to ensure better health even then. To prevent further damage to their health, they find it optimal to call themselves sick and get treated in the current period and then secure improved health for their future selves. On the contrary, we argue it is possible that women with low agency do not tag themselves as sick in current period and hence can potentially compromise their future health. Therefore, our unexpected finding can plausibly indicate that agency enhance private health subtly but counter-intuitively.

## 6. DISCUSSION AND CONCLUSION

The implications of women agency on development outcomes including mental and physical health depends to a large extent on context. Among many probable dimensions, the importance of context on the issue of female agency lies in how gendered the social norms and beliefs are. As an instance, Richardson et al. (2019) documents women agency to be linked with better mental health in parts of rural Rajasthan in India. Nevertheless, they also find women, who are comfortable in speaking up in public suffer from higher mental distress. In this paper, we complement this work using data from the highly urbanized setting of the city of Dhaka. The context here assumes importance as the social gender norms are different in Dhaka in comparison to rural Rajasthan.

The findings from the paper indicate female with higher agency are less likely to suffer from mental distress and depression. This inference is especially driven by those who are comfortable in speaking up in public. Nevertheless, we find counter intuitive results when we examine the implications of agency on physical health of women. In particular, we observe agency to be negatively correlated on an average with the likelihood of being ill even after controlling for the possible confounders including differential time use patterns for women with varying levels of agency. Notably, we are able to replicate these inferences using days of illness as another, to sum up outcome variable. Hence our findings reveal high agency women are likely to be less depressed mentally, yet they are able to enjoy fewer healthy days.

We hypothesize and propose a resolution for this puzzle in terms of purposeful action and ultimately sustainable development: women with higher agency place higher weight on future health. At the heart of women's agency is purposeful action since women with high agency are capable of conceiving purpose in the future and formulating action that is feasible today. Evidently, they respond affirmatively to the statement about speaking up in public to help decide on services needed for their communities. Such ability in these women is likely to prevent them from succumbing to depression. This explains the left arrow in the puzzle (Figure 10).

[Figure 10 here]

The same ability of conceiving purpose for the future and formulating action today drives the right arrow. Precisely because they have a clearer picture of purpose in their future, and they are able to formulate action today which is consistent with such purpose, these women may not hesitate to call in sick today for ensuring that they maintain their health in some definite future. Because of high agency, meaning clearer purpose in the future and consistent action today, the right arrow is necessary.

This resolution to the puzzle can illuminate our understanding of sustainable development and characterise the critical role of women's empowerment and agency. If sustainable development is indeed an expansion of freedoms which simultaneously guarantees the future generation's access to resources for their endeavour, then in individual terms women's agency is essential. On the basis of our findings and possible inferences uncovered here, enhancing women's agency becomes pertinent to secure sustainability.

**FIGURE 1** Location of the surveyed households

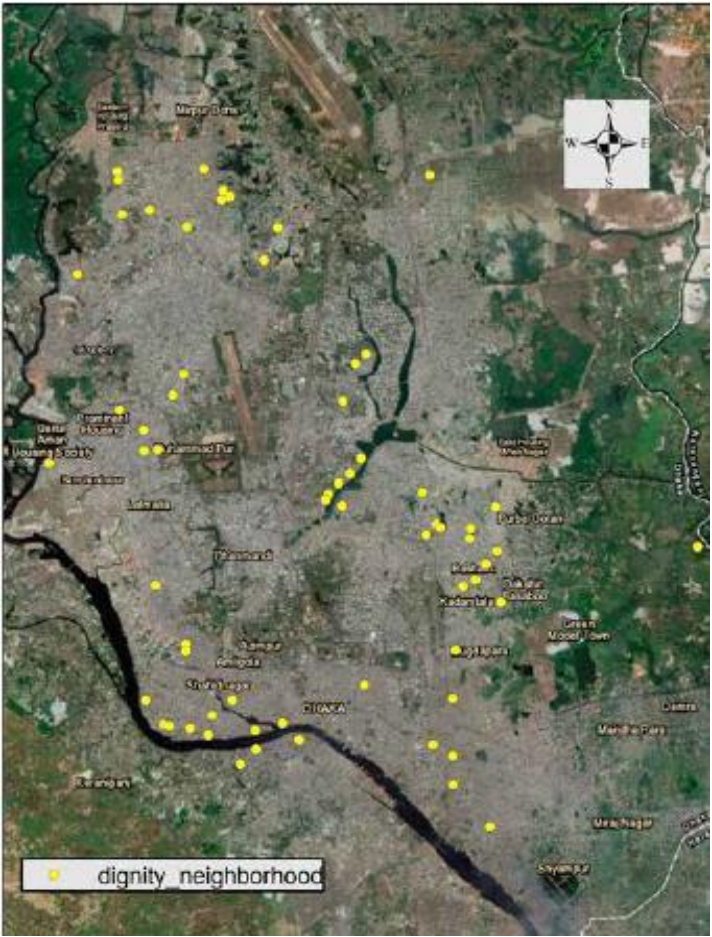

World Bank DIGNITY Survey, 2018

**FIGURE 2** Local polynomial plot of standardized mental health score

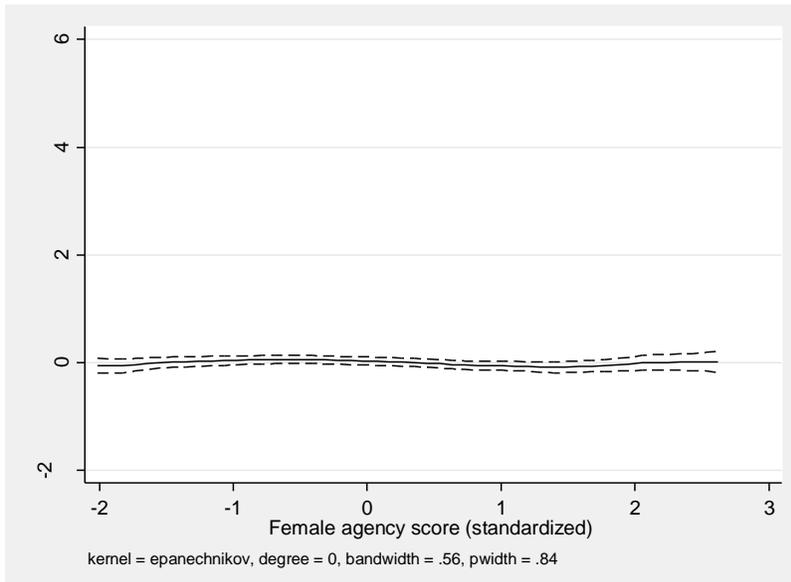

The command "lpoly" in STATA 14 is used to generate the plot. 95% confidence interval is also plotted.

**FIGURE 3** Local polynomial plot of probability of reporting illness

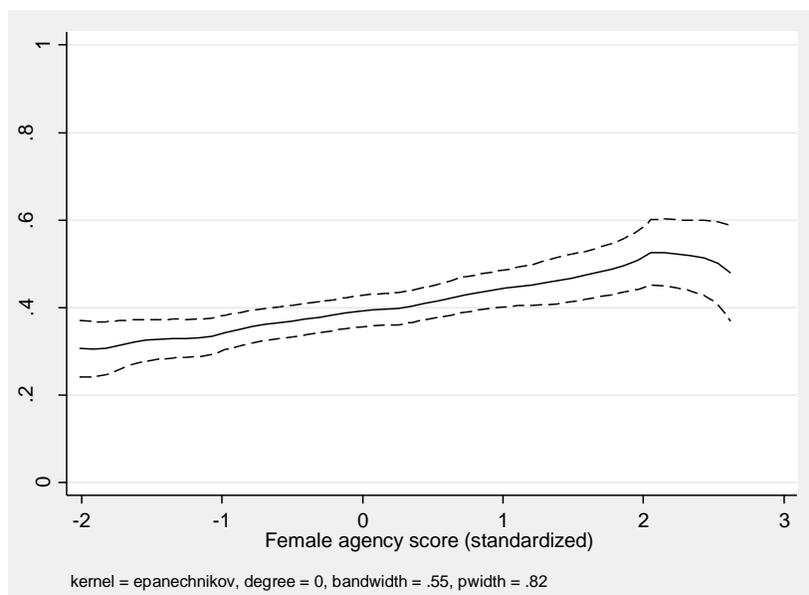

The command "lpoly" in STATA 14 is used to generate the plot. 95% confidence interval is also plotted.

**FIGURE 4** Local polynomial plot of days of illness

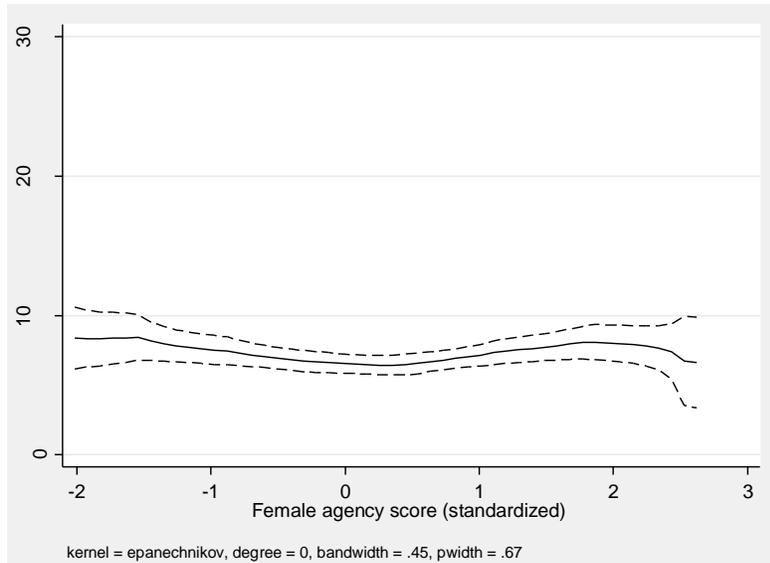

The command "lpoly" in STATA 14 is used to generate the plot. 95% confidence interval is also plotted.

**FIGURE 5** Estimates of indicators of women agency on mental distress.

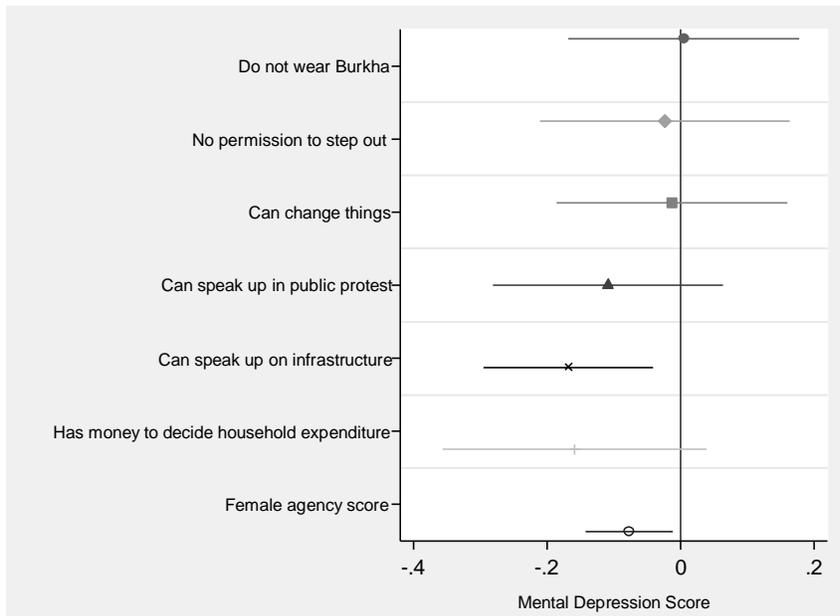

Marginal effects from OLS regressions along with the standard errors clustered at neighbourhood level from which the 95% confidence intervals are calculated have been plotted in the graph. The unit of analysis is females above the age of 15 years. Full regression table is given in online supplementary table A1

**FIGURE 6** Estimates (Odds ratio) of indicators of women agency on the likelihood of being physical ill.

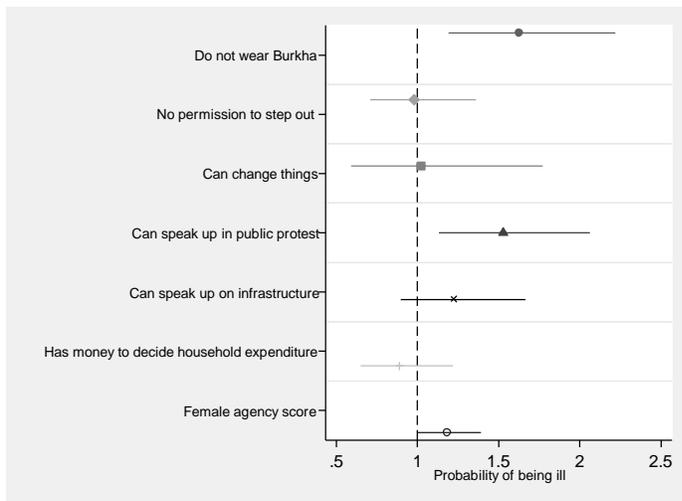

Odds ratio from probit regressions along with the standard errors clustered at neighbourhood level from which the 95% confidence intervals are calculated have been plotted in the graph. The unit of analysis is females above the age of 15 years. Full regression table is given in online supplementary table A2.

**FIGURE 7** Estimates of indicators of women agency on the number of days of being physical ill.

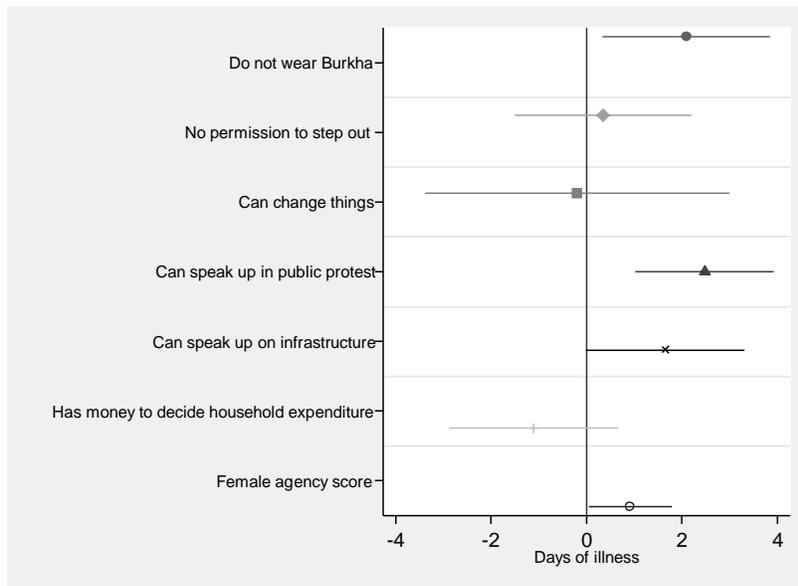

Marginal effects from tobit regressions along with the standard errors clustered at neighborhood level from which the 95% confidence intervals are calculated have been plotted in the graph. The unit of analysis is females above the age of 15 years. Full regression table is given in online supplementary table A3.

**FIGURE 8** Local polynomial plots for time use patterns

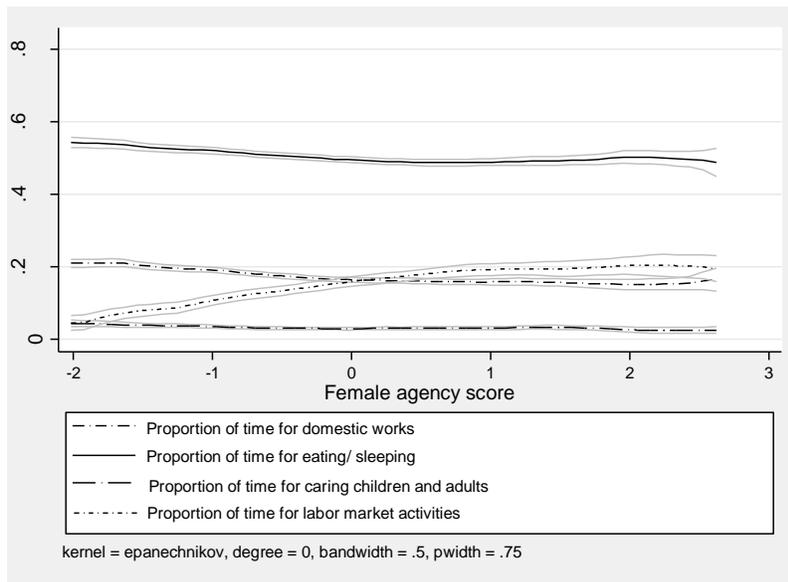

The command "lpoly" in STATA 14 is used to generate the plot. 95% confidence interval is also plotted.

**FIGURE 9** Estimates for being chronic and hospitalized

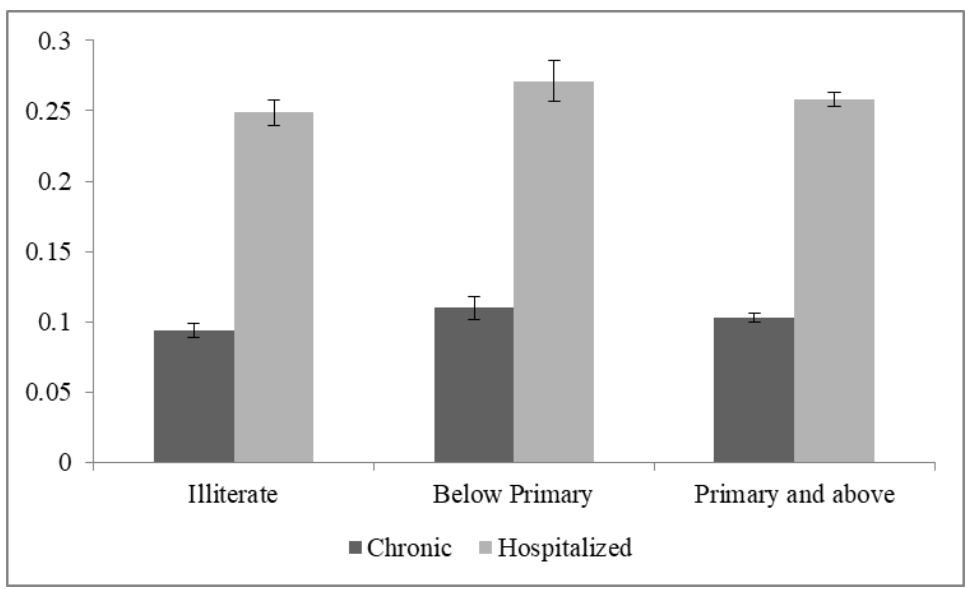

Authors own calculation from NSS 75$^{th}$ round survey. Marginal effects from probit regressions are plotted with the 95% confidence interval clustered at the district level.

**FIGURE 10** The puzzle of women agency, depression and physical health

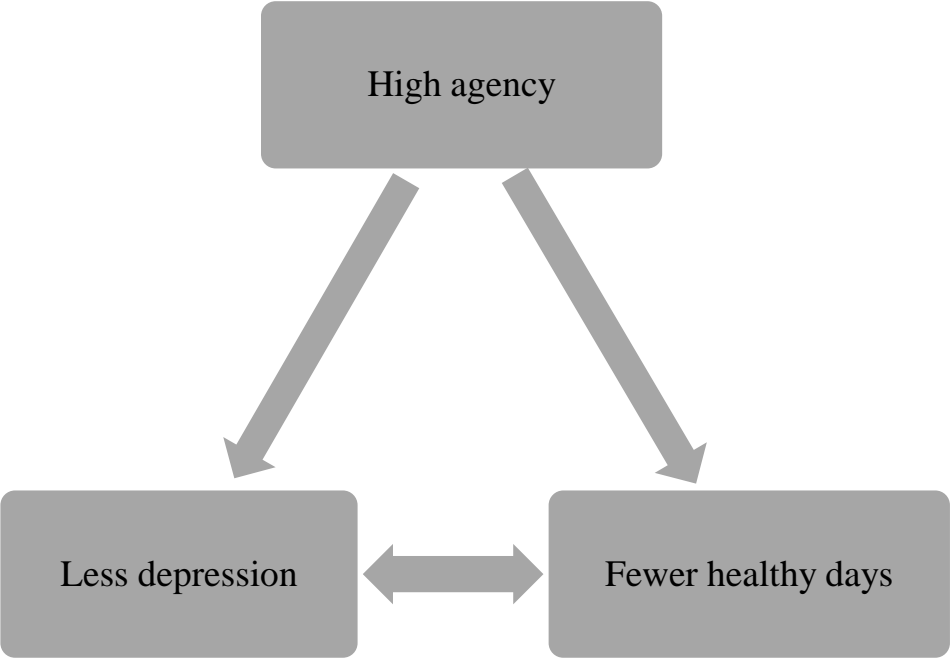

**TABLE 1** Descriptive Statistics

| Variables | Mean | Standard Deviation | Maximum | Minimum |
|---|---|---|---|---|
| *Outcome variables (Continuous)* | | | | |
| Disturbed by things that don't normally bother you | 1.436 | 2.019 | 7 | 0 |
| Faced trouble concentrating on what you are doing | 1.336 | 1.811 | 7 | 0 |
| Feel depressed | 1.579 | 2.055 | 7 | 0 |
| Feel everything that you did was a burden | 1.348 | 2.050 | 7 | 0 |
| Feel afraid | 0.474 | 1.185 | 7 | 0 |
| Restless sleep | 1.222 | 1.799 | 7 | 0 |
| Feel lonely | 0.995 | 1.881 | 7 | 0 |
| Feel like not getting up in the morning | 2.992 | 2.777 | 7 | 0 |
| Days of illness (among those who reported illness) | 7.178 | 6.069 | 30 | 1 |
| *Independent variables (continuous)* | | | | |
| Age of the respondent | 34.114 | 11.863 | 97 | 15 |
| Total consumption expenditure on milk product, egg, fish, meat and milk (in Bangladeshi Taka) | 606.587 | 543.162 | 5915 | 2 |
| Household size | 3.918 | 1.648 | 13 | 1 |
| Number of mobile phones | 1.771 | 0.946 | 6 | 0 |
| *Outcome variables (Binary)* | | | | |
| Variables | | | Proportion | |
| Suffered from illness in the last 30 days | | | 0.393 | |
| *Independent variables (Binary)* | | | | |
| Do not wear burkha | | | 0.514 | |
| No need for permission to step out | | | 0.602 | |
| Can change things | | | 0.099 | |
| Can speak up in public protest | | | 0.357 | |
| Can speak up on infrastructure | | | 0.432 | |
| Has money to decide household expenditure | | | 0.769 | |
| Attended education | | | 0.620 | |

| | |
|---|---|
| Currently married | 0.849 |
| Working outside | 0.524 |
| Respondent is the household head | 0.198 |
| Female headed household | 0.212 |
| Cooking fuel: Liquefied Petroleum Gas (LPG) | 0.748 |
| Cemented roof | 0.278 |
| House is owned/ Rent free | 0.210 |
| Mother is literate | 0.170 |
| Father is literate | 0.289 |
| Brought assets from parents home | 0.494 |
| Observations | 1,086 |